\documentstyle[prl, aps, amssymb
]{revtex}
\begin{document}
\draft
\wideabs{
\title{$^{17}$O NMR study of $q=0$ spin excitations
in a nearly ideal $S=\frac{1}{2}$ 1D Heisenberg antiferromagnet,
Sr$_{2}$CuO$_{3}$, up to 800 K}
\author{K.R. Thurber,$^{1,2}$\cite{address} A.W. Hunt,$^{1,2}$
 T. Imai,$^{1,2}$ and
F.C. Chou$^{2}$}
\address{Department of Physics$^{1}$ and Center for
Materials Science and
Engineering,$^{2}$ M.I.T., Cambridge, MA 02139}
\date{\today}
\maketitle
\begin{abstract}
We used $^{17}$O NMR to probe the uniform (wavevector $q=0$)
electron spin
excitations up to 800 K in Sr$_{2}$CuO$_{3}$
and
separate the $q=0$ from the $q=\pm\frac{\pi}{a}$
staggered components.
Our results support the logarithmic
decrease of the uniform spin susceptibility below
$T \sim 0.015J$,
where $J=2200$ K.  From measurement of the dynamical spin
susceptibility for $q=0$ by the spin-lattice
relaxation rate $1/T_{1}$, we demonstrate that the $q=0$
mode of spin transport is ballistic at the $T=0$ limit, but
has a diffusion-like contribution at finite temperatures even
for $T \ll J$.
\end{abstract}
 } The one-dimensional Heisenberg spin
chain has one of the simplest Hamiltonians, $H = J \sum_{i} {\bf
S}_{i} \cdot {\bf S}_{i+1}$, yet our understanding of its
fascinating quantum mechanical properties is still developing with
recent
theoretical\cite{Affleck,Sachdev,Sandvik,Starykh,Starykh2,Fabricius,Dtheory,Sachdev2}
and experimental\cite{Motoyama,Takigawa1,Takigawa2,Takigawa3}
studies.
% One recent theoretical question has been whether spin
% diffusion occurs
% in a 1D $S=\frac{1}{2}$ Heisenberg
% antiferromagnet\cite{Fabricius,Dtheory,Starykh2}.
A recent breakthrough in experimental
studies of
spin chains is the identification of a nearly ideal
1D $S=\frac{1}{2}$ Heisenberg
antiferromagnet, Sr$_{2}$CuO$_{3}$, by Motoyama {\it et al.}
\cite{Motoyama}  In this
system, $S=\frac{1}{2}$ spins reside at Cu sites, and
the superexchange interaction $J$ is mediated
by hybridization with the $2p_{\sigma}$ orbital of the
in-chain O(1)
site, see Fig.1(a).  Based on the fit of the uniform spin
susceptibility $\chi'(q=0)$ \cite{Affleck}, $J$ is
estimated to be $J=2200\pm 200$ K\cite{Motoyama}.

Sr$_{2}$CuO$_{3}$ has proven to be an ideal material for the
experimental studies of $S=\frac{1}{2}$ Heisenberg spin chain for
various reasons.  First and foremost, weak inter-chain couplings
make the temperature of the N\'eel transition
($T_{N}=5$K)\cite{NeelTemp} to a three-dimensional long-range
ordered state three orders of magnitude smaller than $J$,
$T_{N}=0.002J$.  Thus, the spin excitations of the $S=\frac{1}{2}$
Heisenberg spin chain can be probed at unprecedently low scales of
temperature and energy.  The second major advantage of
Sr$_{2}$CuO$_{3}$ is that $^{63}$Cu NMR is observable at the
magnetic cation site, because the large $J$ suppresses the nuclear
relaxation rates.  In a series of publications, Takigawa {\it et
al.} reported detailed $^{63}$Cu NMR investigations of the
low energy spin excitations\cite{Takigawa1,Takigawa2,Takigawa3}.
They successfully tested the theoretical predictions for the
$q=\pm\frac{\pi}{a}$ staggered mode in the scaling
limit\cite{Sachdev}, including the low temperature logarithmic
corrections to the staggered dynamical
susceptibility\cite{Starykh}.  The third major advantage of
Sr$_{2}$CuO$_{3}$, although it has never been exploited
 in the earlier NMR works, is the high local symmetry
of the crystal structure.  The Cu-O-Cu chain is strictly straight
and the in-chain O(1) site is located in the middle of adjacent Cu
sites as shown in Fig.1(a).  Therefore the staggered components of
the magnetic hyperfine fields from Cu electron spins are canceled
out at the in-chain O(1) sites. Accordingly, one can probe the low
energy spin excitations for the $q=0$ long wavelength mode (see
Fig.1(c)) separately from the staggered $q=\pm\frac{\pi}{a}$ mode.
% We note that the relatively low spectral
% weight of the dynamical structure factor $S(q,\omega)$ near $q=0$
% makes accurate neutron scattering measurement of the $q=0$ mode
% very difficult, if not impossible.
Thus $^{17}$O NMR study of
Sr$_{2}$CuO$_{3}$ provides us with a unique opportunity to
investigate the $q=0$ spin excitations down to $T\sim 0.002J$ in
the nearly ideal $S=\frac{1}{2}$ Heisenberg spin chain without
interference by the staggered mode.
 Unfortunately,
the low oxygen diffusion rate in Sr$_{2}$CuO$_{3}$
severely limits the
$^{17}$O isotope enrichment rate, hence the $^{17}$O
NMR signal intensity.  As such,
no $^{17}$O NMR studies have been reported despite
the rich information expected for the unexplored $q=0$
mode.

In this Letter, we report the first successful
$^{17}$O NMR investigation of
Sr$_{2}$CuO$_{3}$ single crystals.  We accurately
measured the temperature dependence of the uniform
spin susceptibility $\chi'(q=0)$ by NMR Knight shift
at the
in-chain O(1) site {\it without suffering
from the
contribution by free-spins} that limits the accuracy
of bulk susceptibility measurements and
$^{63}$Cu NMR at low temperatures.
We found that $\chi'(q=0)$ decreases steeply below
$T \sim 0.015J$
 without the signature of three-dimensional short range
order
 approaching the N\'eel state.  Our observation
supports the presence of a logarithmic decrease of
$\chi'(q=0)$ at low temperatures, but the quantitative
agreement with existing theoretical models
\cite{Affleck} is not
good.  Based on $^{17}$O NMR spin-lattice relaxation rate
$1/T_{1}$, we also test whether the $q=0$ mode of spin
transport is ballistic or diffusive in a $S=\frac{1}{2}$
1D Heisenberg chain, a long standing
controversy\cite{Sachdev,Fabricius,Dtheory}.  Our data
strongly support the ballistic nature of spin
transport at the $T=0$ limit, but suggest the presence of
diffusion at finite temperatures.

A single crystal of Sr$_{2}$CuO$_{3}$ was grown in a
traveling-solvent
floating-zone furnace.  $^{17}$O isotope was enriched
into crystals
by annealing them at 900-1045 C in $^{17}$O$_{2}$ gas.
 NMR measurements
were conducted by home made NMR spectrometers operated typically
at 9 Tesla.
% Because of the low $^{17}$O NMR intensity in this
% material, acquisition of a data point presented in this Letter
% required typically one to several days of continuous signal
% averaging.
The hyperfine interaction form factor, $F$, between
$^{17}$O nuclear spins and Cu electron spins
 is represented as $F_{1}(q) = 2C cos(\frac{qa}{2})$ and $F_{2}(q) = D$
at the O(1) and O(2) sites, respectively\cite{Moriya}.
$C$ and $D$
represent the hyperfine coupling constant tensor
between the observed
$^{17}$O nuclear spin and the nearest neighbor Cu
electron spins, as
schematically shown in Fig.1(a).  We determined the
hyperfine
coupling tensor based on the standard $K-\chi$ plot
analysis as
$2C^{a}=45$, $2C^{b}=95$, $2C^{c}=44$, $D^{a}=75$,
$D^{b}=23$, $D^{c}=14 \pm 10$ kOe/$\mu_{B}$.  The superscripts represent
the crystal axes.  One kOe/$\mu_{B}$ is a hyperfine interaction of
$g \gamma_{n} \hbar = 4.74 \times 10^{-9}$ eV for $^{17}$O.
% The hyperfine form factor at the O(1) site takes zero
% value
% $F_{1}(q=\pm\pi/a)=0$ at the zone edge $q=\pm\pi/a$.
% This means that $^{17}$O NMR
% properties at the in-chain O(1) sites depend very
% little on the staggerred
%% spin susceptibility, and reflects the properties
% at another low frequency branch of the spectrum near
% $q=0$.
%% In what follows, we utilize this property to probe
% the temperature dependence
% of the uniform spin susceptibility $\chi'(q=0)$ and
% the $q=0$ mode of low
% frequency spin fluctuations through NMR Knight
% shifts $K(1)$ and $1/T_{1}(1)$, respectively.

In Fig.2, we present the temperature dependence of the
NMR Knight shift $K^{a}(1)$
measured at the in-chain O(1) site with 9 Tesla of uniform
magnetic
field applied along the a-axis.  We emphasize that one
cannot achieve
such extremely high
experimental accuracy at the
O(2) and Cu sites in low temperatures, because the
enhanced staggered spin susceptibility near
$q=\pm\pi/a$ causes both homogeneous line
broadening as well
as inhomogeneous line broadening arising
from defects\cite{Takigawa3}.  In contrast
$F_{1}(q=\pm\frac{\pi}{a})=0$ implies
that these line-broadening mechanisms are ineffective
at the in-chain O(1)
sites, and the linewidth remains one to two orders of
magnitude
narrower than at the O(2) and Cu sites.  The O(1)
linewidth is only 8 kHz at 30 K, compared to 50 kHz for the O(2) linewidth.  The
NMR shift $K^{a}(1)$ can be expressed as
\begin{equation}
K^{a}(1)=\frac{2C^{a}}{N_{A}\mu_{B}}\chi'(q=0) +
K_{VV}^{a}(1).
\label{K}
\end{equation}
where the first term ($K^{a}_{spin}(1)$ in Fig.2) arises from the uniform spin
susceptibility
$\chi'(q=0)$, while the second term
$K_{VV}^{a}(1)=0.024\%$ represents
the small temperature independent contribution of the
Van-Vleck term as determined by the $K-\chi$ plot
analysis.
The temperature dependence above 30 K of
$\chi'(q=0)$ can be fitted very well with the high
temperature
theory\cite{KnhighT,Motoyama}.  Below 30 K ($\sim
0.015J$), $\chi'(q=0)$ begins to
decrease steeply.  We may compare the steep decrease of
$\chi'(q=0)$
 with the theoretical prediction by
 Eggert, Affleck, and Takahashi\cite{Affleck} based on
conformal field theory.  They
predicted a logarithmic term which has a steep
 decrease with an infinite slope at the zero temperature
limit, $J\pi^{2}\chi'(q=0)\sim
 1+\frac{1}{2ln(T_{o}/T)}$ with $T_{o}\sim 7.7J$, as
shown in the inset to Figure 2.  At the qualitative level, our data support the
presence of a steep decrease of $\chi'(q=0)$. However, our
results show that the theoretically predicted logarithmic term in
$\chi'(q=0)$ does not account for our experimental results at the
quantitative level for $T \lesssim 15$ K.
We note that Motoyama {\it et al.}\cite{Motoyama}
earlier concluded that their bulk susceptibility
measurements matched the theoretical result above $\sim 10$ K.
However, we emphasize that our
$^{17}$O NMR Knight shift data do not suffer from the Curie term
observed in the bulk susceptibility data.  As already pointed out
by Motoyama {\it et al.}\cite{Motoyama}, the necessity of subtracting
the Curie term $\sim1/T$ with a divergently large {\it negative}
temperature coefficient from the bulk susceptibility data leaves
ambiguities in the temperature dependence of the potential
logarithmic term deduced after the subtraction, because the latter
has a divergently large {\it positive} temperature coefficient. In
contrast, $K^{a}(1)$ directly reflects the local spin
susceptibility $\chi'(q=0)$ of the neighboring Cu electron spins
and has no such ambiguity.  Thus the results in Fig.2 provide an
unambiguous experimental test for theories.

Furthermore, we can rule out the
possibility\cite{Motoyama} that the
steep decrease
of $\chi'(q=0)$ and/or poor agreement with theory
 is caused by the three-dimensional short-range order
near the
N\'eel transition at $T_{N}=5 K$ based on the
following
consideration.  The temperature dependence of the
$^{17}$O nuclear
spin-lattice relaxation rate $1/T_{1}^{a}(2)$ at the
O(2) site is presented
in Fig.3(a).  $1/T_{1}^{a}$ at
O(1) and O(2) sites may be expressed as
\begin{equation}
          \frac{1}{T_{1}^{a}} =
\frac{\gamma_{n}^{2}k_{B}T}{\mu_{B}^{2}}
    \sum_{q} [|F^{b}(q)|^{2}+|F^{c}(q)|^{2}]
\frac{\chi''(q,\omega_{n})}{\omega_{n}}
\label{T1}
\end{equation}
 where $\gamma_{n}$ is the nuclear gyromagnetic ratio
of the observed
nucleus ($\gamma_{n}=5.772$ MHz/Tesla for $^{17}$O
nuclei), and
$\omega_{n}$($\sim\gamma_{n}H$) is the resonance
frequency.
Since the form factor at the O(2) site $F_{2}(q)$ is
independent of wavevector $q$, $1/T_{1}(2)$ at the
O(2) sites measures the
wavevector $q$ integral of
$\chi''(q,\omega_{n})$.  This represents
 the strength of the low frequency ($\omega_{n}$)
Cu electron spin fluctuations.  The roughly constant value of
$1/T_{1}(2)$ reflects the cancellation of the two temperature
dependent parts of eq.\ref{T1}.  $1/T_{1} \propto T$ which offsets the
$\sim 1/T$ increase of the q-integral of the staggered spin susceptibility
$\chi''(q=\pm\frac{\pi}{a},\omega_{n})$ down to 10 K,
 as previously reported by Takigawa {\it et
al.}\cite{Takigawa1} and Thurber {\it et al.}\cite{3-leg} based on
$^{63}$Cu NMR.  We note that the mild increase of $1/T_{1}(2)$
down to 10 K is consistent with the logarithmic correction
intrinsic to 1D behavior\cite{Takigawa1}.  Only at the lowest
temperature measured (4.2 K)
does $1/T_{1}(2)$ strongly increase, indicating the
three-dimensional order of the N\'eel state at
$T_{N}=5$ K.  We emphasize that the effect of three-dimensional
short range order is
not seen in $1/T_{1}(2)$ for the region 30 K down to 10 K, where a
steep decrease of
$\chi'(q=0)$ is evident in our $K^{a}(1)$ data.  This implies that
the decrease of $\chi'(q=0)$ is a property of the 1D spin chains,
not a 3D ordering effect.  On the other hand, we cannot rule out
the possibility that the discrepancy between theory and experiment
is caused by the finite length of the spin chains caused by defects.
The
finite chain length produces a staggered ($q=\pm\frac{\pi}{a}$) spin
density oscillation in this temperature regime \cite{Takigawa3}
which might cause unknown effects to $\chi'(q=0)$.

Next we turn our attention to the temperature
dependence of the
long-wavelength $q=0$ mode of the low
frequency Cu electron spin fluctuations.
The temperature dependence of $1/T_{1}^{a}T$ at the
in-chain O(1) and apical O(2) sites is compared in Fig.4(a).
We found that $1/T_{1}^{a}(1)T$ may be approximated by
an empirical form,
$1/T_{1}^{a}(1)T=0.027 + 4.7\times10^{-4}T$
sec$^{-1}$K$^{-1}$ at low
temperatures.  $1/T_{1}^{a}(1)T$ shows qualitatively
different behavior
from $1/T_{1}^{a}(2)T$, because the hyperfine form
factor
$F_{1}(q=\pm\frac{\pi}{a})=0$ {\it filters out} the
contribution of the staggered
 susceptibility  and $1/T_{1}^{a}(1)T$ is
dominated by  $q \sim 0$ (see Fig.1(c)).  According to the
theoretical prediction by Sachdev\cite{Sachdev}
based on quantum critical scaling at the low temperature
limit, the
$q=0$ contribution to  $1/T_{1}^{a}(1)T$ may be
written as
$1/T_{1}^{a,q=0}T=[(2C^{b})^{2}+(2C^{c})^{2}]g^{2}\gamma_{n}^{2}\hbar
k_{B}/\pi^{3}J^{2}$.
The underlying assumption is that the $q=0$ spin
excitations propagate ballistically without damping at low
temperatures rather than diffusive transport.
By inserting $C^{b}$, $C^{c}$, and $J$ into the
scaling form,
we obtain the theoretical estimate of the contribution
by the undamped
ballistic mode,
$1/T_{1}^{a,q=0}(1)T=0.029\pm0.006$ sec$^{-1}$K$^{-1}$.
This is in excellent
agreement with our experimental zero temperature
limit, $1/T_{1}^{a}(1)T=0.027\pm 0.004$ sec$^{-1}$K$^{-1}$
without any adjustable parameters.  We note that
the
contribution by the
$q=\pm\frac{\pi}{a}$ branch,
$1/T_{1}^{q=\pm\frac{\pi}{a}}(1)T$ is not
negligible at $T\neq 0$, because the
form factor $F_{1}(q)$ will be finite for any
$q\neq\pm\frac{\pi}{a}$.  However by integrating the
wave-vector
dependent scaling form of the staggered
susceptibility\cite{Sachdev} convoluted by the form
factor
$|F_{1}(q)|^{2}$,
we estimate $1/T_{1}^{q=\pm\frac{\pi}{a}}(1)T$ as only
$\sim$5\% of the observed
rate at 300 K, and the percentage contribution decreases
slowly with decreasing temperature.
Thus the scaling
estimation of the sum of the two separate modes of
contributions,
$1/T_{1}^{q=0}(1)T+1/T_{1}^{q=\pm\frac{\pi}{a}}(1)T$,
while providing a very good estimate for $T=0$,
severely underestimates our
experimental results {\it at any finite temperature},
as shown by a dashed line in Fig.4.  Moreover, we found
that $1/T_{1}(1)$ depends on frequency as shown in Fig. 3(c).

Thus, there is an additional contribution to the low energy spin
susceptibility for $q \sim 0$ that increases strongly with
increasing temperature and decreasing frequency. This suggests
that spin diffusion\cite{DeGennes} is important, even for $T \ll
J$.  We measured the frequency dependence of $1/T_{1}(1)$ between
$H=7$ Tesla ($\omega_{n}=\gamma_{n}H=40.4$ MHz) and $H=14$ Tesla
($\omega_{n}=80.8$ MHz) at 77 K and 295 K. The mild frequency
dependence of $1/T_{1}(1)$ presented in Fig.3(c) is consistent
with the $1/\sqrt{\omega_{n}}$ dependence\cite{Borsa} expected for
the diffusive contribution,
\begin{equation}
          \frac{1}{T_{1}^{diff,a}T} =
[(2C^{b})^{2}+(2C^{c})^{2}]\frac{\gamma_{n}^{2}k_{B}\chi'(q=0)}{2\mu_{B}^{2}
          \sqrt{2\omega_{e}D_{s}}}
\label{T1-diff}
\end{equation}
where $\omega_{e}=g\mu_{B}H/\hbar\propto \omega_{n}$.
The $^{63}$Cu NMR $1/T_{1}$\cite{Takigawa1} also had a small
frequency dependent component consistent with
$1/\sqrt{\omega_{n}}$, even though Cu NMR is dominated
by the $q=\pm\frac{\pi}{a}$ modes.  $1/T_{1}(2)T$ for the
O(2) sites does not
have any frequency dependence within error, indicating
that the dominant
$q=\pm\frac{\pi}{a}$ modes are not diffusive.
To further
establish the presence of an unexpectedly large
diffusive contribution
in the $S=\frac{1}{2}$ Heisenberg spin chain,
we also measured $1/T_{1}(1)$ in a
related one-dimensional spin chain system SrCuO$_{2}$
(see Fig.1(b)
and Fig.3(b)).  Since the signal intensity
of $^{17}$O NMR is strong in SrCuO$_{2}$ owing to
the higher isotope enrichment rate, we could measure
the frequency
dependence of $1/T_{1}(1)$ with higher accuracy.
Because of
the transferred hyperfine coupling $D' (\sim D)$ from
the adjacent chain,
to a good approximation $1/T_{1}(1)$ in SrCuO$_{2}$
is a superposition of the contributions from the $q=0$
modes and $1/T_{1}(2)$.
This explains why $1/T_{1}(1)$ in SrCuO$_{2}$
asymptotes to
$1/T_{1}(2)$ at the low temperature limit as shown in
Fig.3(b).
The presence of a large contribution with
$1/\sqrt{\omega_{n}}$-dependence
is clearly seen in Fig.3(c).

Theoretically, even whether spin diffusion exists for the $q=0$
mode of the $S=\frac{1}{2}$ 1D Heisenberg spin chain is
controversial\cite{Sachdev,Starykh2,Fabricius,Dtheory,Sachdev2}.
Spin diffusion has been measured by NMR in $S=\frac{5}{2}$ spin
chains\cite{Borsa} and a $S=1$ Haldane-gap system\cite{Takigawa4},
but to our knowledge not for a $S=\frac{1}{2}$ system.  Quantum
Monte Carlo results\cite{Sandvik,Starykh2} indicate a strong
increase (faster than $T$) of $1/T_{1}(q=0)$ with temperature, but
do not determine if the $q=0$ peak is truly diffusive.  Fabricius
and McCoy\cite{Fabricius} have calculated that the frequency
dependence may be $\sim \omega^{-0.3}$ rather than
$\omega^{-0.5}$, while a finite size scaling analysis suggests
ballistic behavior\cite{Dtheory}.  Our $^{17}$O $1/T_{1}$ results
clearly show frequency dependence, but are not accurate enough to
conclusively distinguish the exact exponent. If we assume that the
extra contribution to $1/T_{1}T$ is genuinely diffusive with
frequency dependence $\omega^{-0.5}$, we can estimate the
temperature dependence of the spin diffusion constant $D_{s}$,
which is the only unknown parameter in eq.\ref{T1-diff}.  In such
a scenario, $D_{s} \sim 1/T^{2}$ for $T \ll J$ as shown in
Fig.4(b).  We caution however that this estimate of $D_{s}$ is a
lower bound on the value, since we are assuming that the
additional contribution to $T_{1}$ is entirely diffusive.

To conclude, we have successfully separated the $q=0$
mode in both
static and dynamic spin susceptibility in
a nearly ideal 1D $S=\frac{1}{2}$ Heisenberg antiferromagnet
material.
We unambiguously demonstrated a steep decrease below
$T=0.015J$ of $\chi'(q=0)$.
Measurements of the low frequency $q \sim 0$ dynamic spin
susceptibility have a $T=0$ limit that agrees with purely
ballistic spin transport.  However, with increasing temperature,
the dynamic spin susceptibility strongly increases.  This result
establishes the presence of non-ballistic behavior at finite
temperatures, even for $T \ll J$.
We suggest the increased dynamic spin susceptibility is from
diffusive contributions, and estimated a lower bound on the
diffusion constant $D_{s}\sim 1/T^{2}$.  Whether these
new results can be accounted
for by the one-dimensional $S=\frac{1}{2}$ Heisenberg
model remains to be
seen.

This work was supported by NSF DMR 98-08941.

% now the references. delete or change fake bibitem.
% delete next three
%   lines and directly read in your .bbl file if you
% use bibtex.

% Fill in the caption in the braces of the \caption{}
% command. Put the label
% that you will use with \ref{} command in the braces
% of the \label{} command.
%
% \begin{figure}
% \caption{}
% \label{}
% \end{figure}

\begin{figure}
\caption{The fundamental building block of the Cu-O
spin chain (Cu $\bullet$, O $\circ$) in
(a) Sr$_{2}$CuO$_{3}$, and (b) a related zig-zag
spin chain material  SrCuO$_{2}$.  Arrows
with $C$, $D$, and $D'(\sim D)$ show the path of
transferred hyperfine
interactions.
(c) The spin excitation spectrum of $S=\frac{1}{2}$
1D Heisenberg antiferromagnetic spin chain.}
\label{structurefig}
\end{figure}

\begin{figure}
\caption{$\bullet$: Temperature dependence of the $^{17}$O NMR
Knight shift $K^{a}(1)$ measured at the in-chain O(1) sites in
Sr$_{2}$CuO$_{3}$.  Solid curve represents analytic calculations
of uniform spin susceptibility $\chi'(q=0)$ by
M. Takahashi {\it et al.}
%\cite{KnhighT}
 for $J=2200$ K, $2C^{a}=45$
kOe/$\mu_{B}$, and $K_{VV}^{a}=0.024$\%. Inset: The same data in
an extended scale. The solid curve is the theoretically predicted
logarithmic form.
%\cite{Affleck}
} \label{Kfig}
\end{figure}

\begin{figure}
\caption{(a) and (b): $1/T_{1}^{a}(2)$ in Sr$_{2}$CuO$_{3}$
($\bullet$), $1/T_{1}^{b}(2)$ in SrCuO$_{2}$ ($\triangle$), and
$1/T_{1}^{b}(1)$ in SrCuO$_{2}$ ($\times$). (c) Frequency
dependence of $1/T_{1}^{a}(1)T$ in Sr$_{2}$CuO$_{3}$ at 77K
($\square$) and 295 K ($\circ$), $1/T_{1}^{a}(2)T$ in
Sr$_{2}$CuO$_{3}$ at 295 K ($\bullet$), and ($1/T_{1}^{b}(1)T -
1/T_{1}^{b}(2)T$) in SrCuO$_{2}$ at 295 K (X). The
$1/\sqrt{H}=1/\sqrt{\omega_{n}}=0$ limit is the theoretical
estimate of $1/T_{1}^{q=0}(1)T+1/T_{1}^{q=\pm\frac{\pi}{a}}(1)T$.
Solid lines are the best linear fit with the theoretical
constraint at $1/\sqrt{H}=0$.  For a contribution from spin
diffusion, the slope is proportional to $1/\sqrt{D_{s}}$.}
\label{1/T1}
\end{figure}

\begin{figure}
\caption{(a)$1/T_{1}^{a}T$ for in-chain O(1) ($\bullet$), and O(2)
($\circ$), and $1/T_{1}^{c}T$ for $^{63}$Cu ($\diamond$, divided
by a factor of 112) in Sr$_{2}$CuO$_{3}$.  The solid line is the
best empirical fit to O(1), $1/T_{1}^{a}(1)T=0.027+4.7\times
10^{-4}T$ sec$^{-1}$K$^{-1}$.  The dashed line represents the
scaling limit estimate of 
$1/T_{1}^{a,q=0}(1)T+1/T_{1}^{a,q=\pm\frac{\pi}{a}}(1)T$. (b)
Temperature dependence of the spin diffusion constant $D_{s}$
deduced from $1/T_{1}^{a}T(1)$.  This deduced value represents a
lower bound on the spin diffusion constant. The high temperature
limit, $D_{s}=(J/\hbar)\sqrt{2\pi S(S+1)/3} = 3.6\times 10^{14}
sec^{-1}$, is also shown by a dashed line. } \label{1/T1T}
\end{figure}

% Fill in the caption in the braces of the \caption{}
% command. Put the label
% that you will use with \ref{} command in the braces
% of the \label{} command.
% Insert the column specifiers (l, r, c, d, etc.) in
% the empty braces of the
% \begin{table}
% \caption{}
% \label{}
% \end{tabular}
% \end{table}

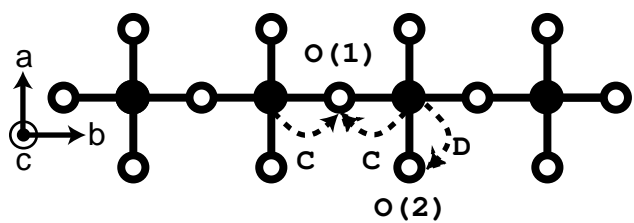
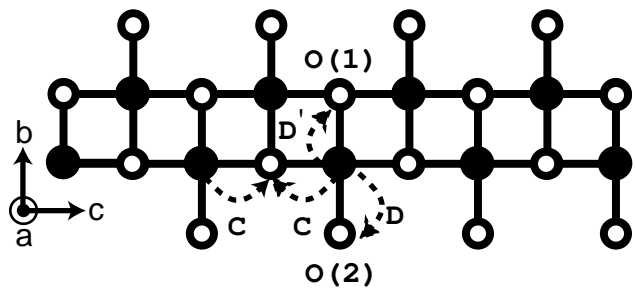
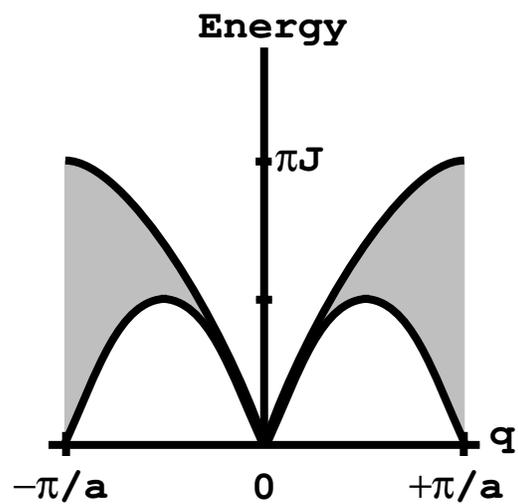
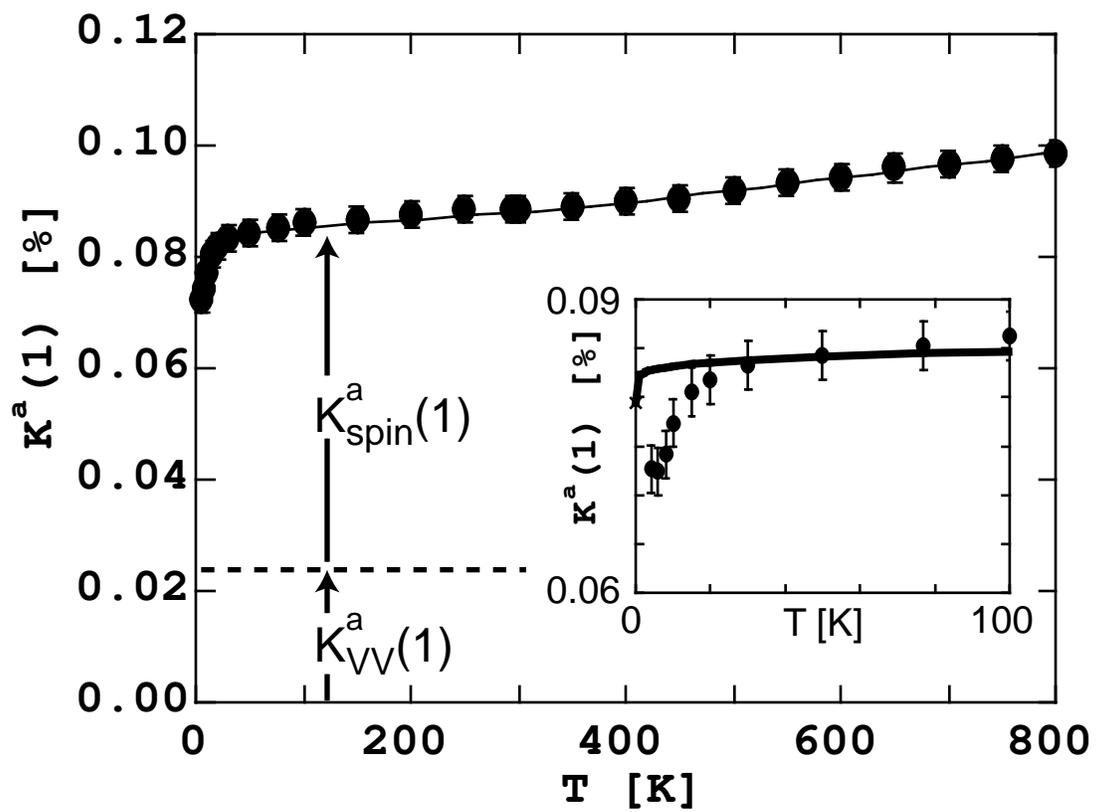
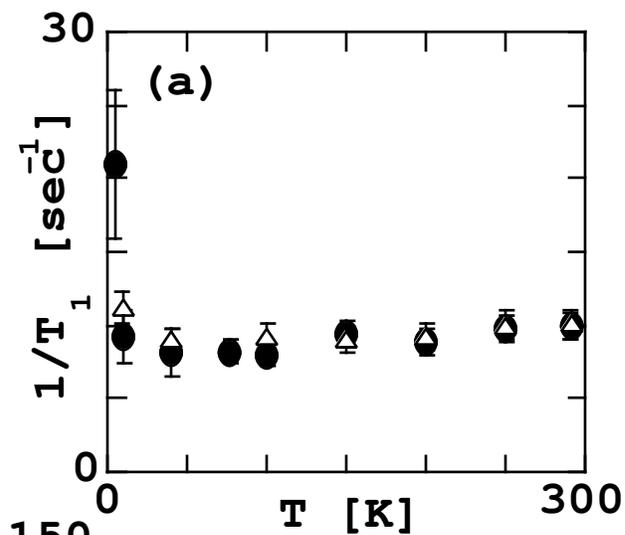
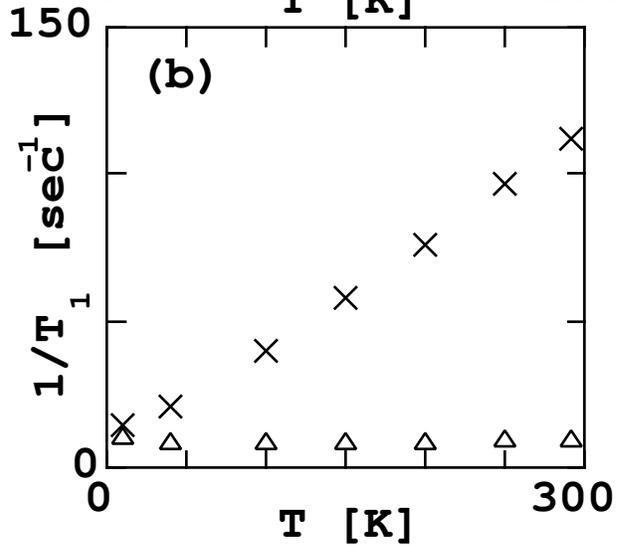
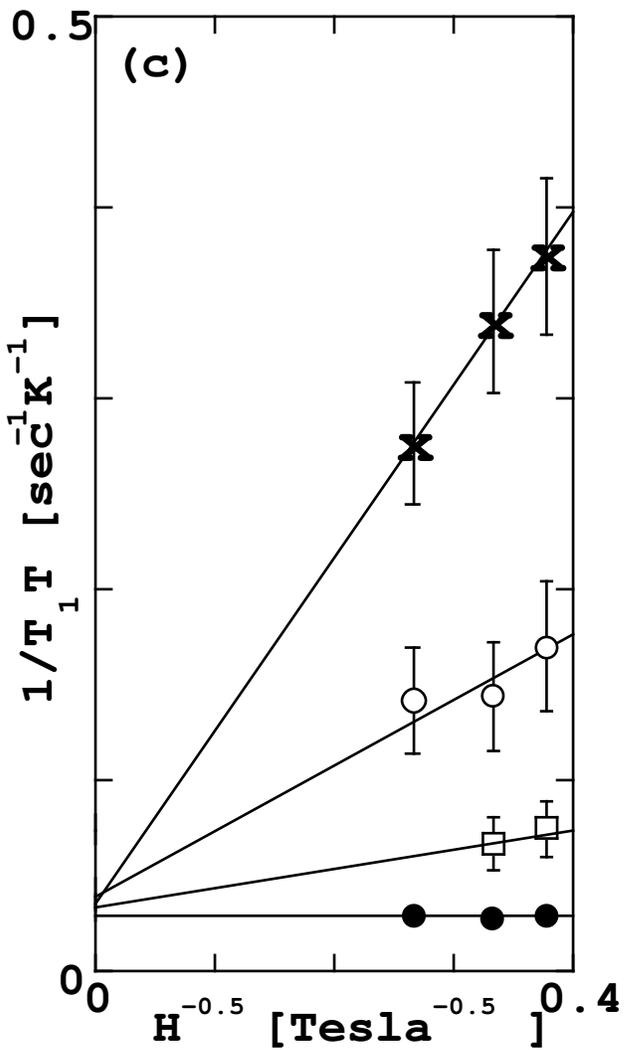
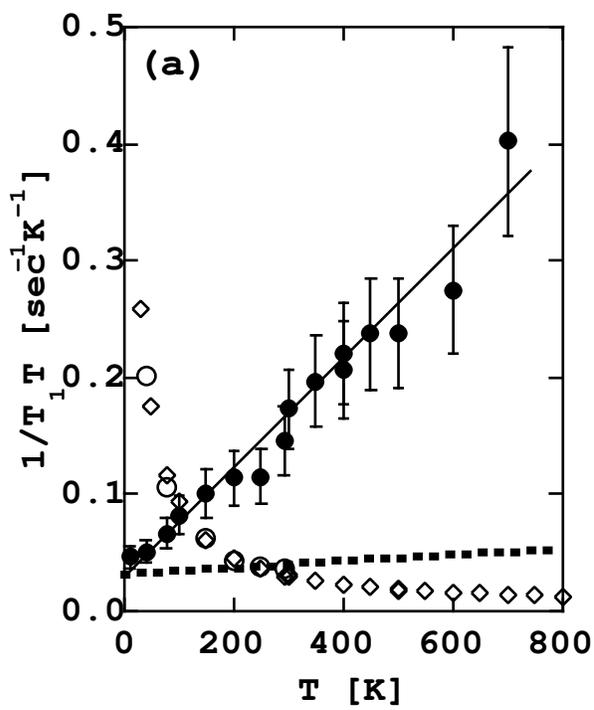
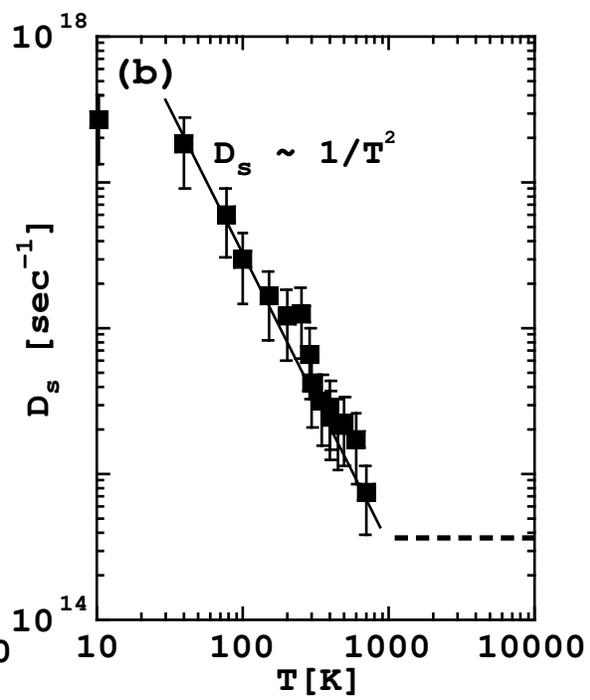
\end{document}